\documentclass[prb,twocolumn,preprintnumbers,amsmath,amssymb]{revtex4}
\usepackage{graphicx}

\begin{document}
\title{ Duality, Magnetic space group and  their  applications to quantum phases and phase transitions
        on bipartite lattices in several experimental systems }
\author{ \bf  Jinwu Ye  }
\affiliation{ Department of Physics, The Pennsylvania State
University, University Park, PA, 16802 }
\date{\today}

\begin{abstract}

    By using a dual vortex method, we study phases such as superfluid, solids, supersolids
    and quantum phase transitions in a unified scheme
    in extended boson Hubbard models at and slightly away from half filling on bipartite optical lattices
    such as honeycomb and square lattice. We also map out its global phase diagram at $ T=0 $
    of chemical potential versus the ratio of kinetic energy over the interaction.
    We stress the importance of the self-consistence condition on
    the saddle point structure of the dual gauge fields in the translational symmetry breaking insulating
    sides, especially in the charge density wave side.
    We find that in the translational symmetry breaking side, different kinds of supersolids are generic
    possible states slightly away from half filling. We propose a new kind of supersolid: valence bond
    supersolid ( VB-SS). In this VB-SS, the density fluctuation at any site is very large
    indicating its superfluid nature, but the boson kinetic energies
    on bonds between two sites are given and break the lattice translational
    symmetries indicating its valence bound nature.
    We show that the quantum phase transitions from solids to
    supersolids driven by a chemical potential are in the
    same universality class as that from a Mott insulator to a
    superfluid, therefore have exact exponents  $ z=2, \nu=1/2, \eta=0 $ with a logarithmic correction.
    Comparisons with previous quantum Monte-Carlo (QMC) simulations on a square lattice are made.
    Implications on possible future QMC simulations in both bipartite lattices  are given.
    All these phases and phase transitions can be potentially realized in
    ultra-cold atoms loaded on optical bipartite lattices.
    Then we apply our results to investigate the reentrant "superfluid"  in a
    narrow region of coverages in the second layer of $^{4}He $
    adsorbed on graphite and  the low temperature phase diagram of
    Hydrogen physisorbed on Krypton-preplated graphite ( $ H_2 $/Kr/graphite ) near half filling.
    We suggest that $^{4}He $ and $ H_2 $ lattice supersolids maybe
    responsible for the experimental signals in the two systems.
    Finally, we suggest Cooper supersolid is repressible for the
    phase diagram of $ La_{2-x} Ba_{x} Cu O_{4} $  near $ x=1/8 $.

\end{abstract}
\vspace{0.2cm}

\maketitle


\section{ Introduction. }

 The Boson Hubbard model with various  kinds of interactions, on all kinds of lattices and at different filling factors
   is described by the following Hamiltonian \cite{boson}:
\begin{eqnarray}
  H & = & -t \sum_{ < ij > } ( b^{\dagger}_{i} b_{j} + h.c. ) - \mu \sum_{i} n_{i} + \frac{U}{2} \sum_{i} n_{i} ( n_{i} -1 )
                                  \nonumber   \\
      & + & V_{1} \sum_{ <ij> } n_{i} n_{j}  + V_{2} \sum_{ <<ik>> } n_{i} n_{k} + \cdots
\label{boson}
\end{eqnarray}
    where $ n_{i} = b^{\dagger}_{i} b_{i} $ is the boson density, $ t $ is the nearest neighbor hopping which
    is determined by the depth of the trapping potential at the prefered adsorption sites.
    $ U, V_{1}, V_{2} $ are onsite, nearest neighbor (nn) and next nearest neighbor (nnn) interactions respectively,
    the $ \cdots $ may include further neighbor interactions
    and possible ring-exchange interactions.

  In the hard-core limit $ U \rightarrow \infty $,
  due to the exact mapping between the boson operator and the spin $ s=1/2 $ operator:
  $ b^{\dagger}_{i} = S^{+}_{i}, b_{i}= S^{-}_{i}, n_{i} = S^{z}_{i} + 1/2 $,
  the boson  model Eqn.\ref{boson} can be mapped to a  "generalized" anisotropic
  $ S= 1/2 $ quantum Heisenberg model in an external magnetic field \cite{vbs,third}:
\begin{eqnarray}
    H & = & - 2 t \sum_{<ij>} ( S^{x}_{i} S^{x}_{j} + S^{y}_{i} S^{y}_{j} )
                         \nonumber   \\
      & + & V_{1} \sum_{<ij>}  S^{z}_{i} S^{z}_{j}
        + V_{2} \sum_{ <<ik>> }  S^{z}_{i} S^{z}_{k} - h \sum_{i} S^{z}_{i} + \cdots
\label{spin}
\end{eqnarray}
    where $ h= \mu - 2  V_{1} - 2 V_{2} $ for a square lattice and the $ \cdots $ may
    include further neighbor interactions and ring-exchange interactions.

     The model Eqn.\ref{boson} with only the onsite interaction was first studied in Ref.\cite{boson}.
     It was found that there is a second order superconductor to insulator transition at filling factor $ f=1
     $. The effects of long range Coulomb interactions on the transition was studied in \cite{coul,yeboson}.
    The duality transformation to the dual vortex picture at $ f=n $ was
    performed in \cite{dualint} and is  briefly reviewed in the appendix.
    Here, we only outline
    its essence to facilitate the duality transformation in the  more
    general cases $ f=p/q $ to be discussed in the next several
    paragraphs.
    In the direct boson picture, a vortex is a singularity in the boson wavefunction, so a boson
    wavefunction acquires a $ 2 \pi $ phase when in encircles a vortex.
    In the dual vortex picture, a boson is a singularity in the vortex wavefunction, so a
    vortex  wavefunction acquires a $ 2 \pi $ phase when in encircles a
    boson. After performing the boson-vortex duality transformation, the authors in \cite{dualint}
    obtained a dual theory of Eqn.\ref{boson}
    in term of the interacting vortices $ \psi $ hopping on the dual lattice subject to a fluctuating
    {\em dual} " magnetic field".
    The average strength of the dual " magnetic field " $ A_{\mu} $
    through a dual plaquette is equal to the boson density $ f=n
    $, because $ 2 \pi n $ is equivalent to $ 0 $, so the average value can be simply taken to be zero.
    It is important to stress that the average density of bosons is the
    same in both the SF and the Mott insulating side, namely, it
    takes the integer $ n $ on both sides, so the average strength of the dual magnetic field can be
    taken as  {\em zero } on both side, then the fluctuations in the dual
    gauge field reflects the fluctuations of the boson density.
    In the continuum limit, just from the gauge invariance,
    the final effective theory in terms of the vortex order parameter $ \psi $
    is just the scaler electrodynamics of the dual vortex $ \psi $
    coupled to the fluctuating gauge field $ A_{\mu} $ described by Eqn.\ref{vortex1}.
    In the superfluid state $  < \psi > =0 $, the gauge field is gapless,
    while in the insulating state $ < \psi > \neq 0 $,
    the gauge field acquires a mass due to the Higgs mechanism.
    This duality transformation has been confirmed by Quantum Monte-
    Carlo (QMC ) in the first reference in \cite{dualint}.

    Obviously, it is important to study other
    commensurate filling factors $ f=p/q $ ( $ p, q $ are relative prime numbers
    ), because  in addition to the superfluid phase, many other phases such as charge density wave
    or valence bond solid can only appear when $ q \geq 2 $. So
    novel phase transitions such as SF to CDW or SF to VBS can be
    realized only in such cases.
    Recently, the dual vortex method ( DVM ) developed in \cite{dualint} was greatly expanded
    to study  Eqn.\ref{boson} on a square lattice
    just at such generic filling factors in \cite{pq1}. It turns out that the DVM at $
    q \geq 2 $ is much more involved than the simplest $ q=1 $ case due to
    the crucial involvement of the non-commutative projective space group at $ q \geq 2 $.
    The general procedures are the following.
    After performing the similar charge-vortex duality transformation as in \cite{dualint},
    the authors in \cite{pq1}  obtained a dual theory of Eqn.\ref{boson}
    in term of the interacting vortices $ \psi_{a} $ hopping on the dual lattice subject to a fluctuating
    {\em dual} " magnetic field".
    The average strength of the dual " magnetic field "  through a dual plaquette is equal to the boson density $ f=p/q $.
    This is similar to the Hofstadter problem
    of electrons moving in a crystal lattice in the presence of a magnetic field \cite{hof,zak}.
    The projective representation of the space group (PSG)
    dictates that there are at least $ q $-fold degenerate minima in the mean field energy spectrum
    ( when $ q=1 $, the PSG just reduces to the usual commutative space group ).
    Near the superconductor to the insulator transition, the most important $ \psi_{a} $ fluctuations will be near these
    $ q $ minima which can be labeled as
    $ \psi_{l}, l=0,1,\cdots, q-1 $ which forms a $ q $ dimensional representation of the PSG.
    In the continuum limit, the final effective theory in terms of these $ q $ order parameters should be
    invariant under this PSG.
    In the superfluid state $  < \psi_{l} > =0 $ for every $ l $, while in the insulating state $ < \psi_{l} > \neq 0 $ for
    at least one $ l $.  In the insulating state, there must exist some kinds of charge density wave (CDW)
    or valence bond solid ( VBS) states which may be stabilized by longer range interactions or possible
    ring exchange interactions included in Eqn.\ref{boson}. The CDW or VBS
    order parameter was constructed to be the most general {\em bilinear} and gauge invariant
    combinations of the $ \psi_{l} $ \cite{pq1}.

    In a recent unpublished preprint \cite{univ}, the author
    studied all the possible phases and phase transitions in
    the EBHM {\em slightly away} from half filling  $ q=2 $ on
    bipartite lattices such as honeycomb and square lattice.
    It was found  in \cite{univ} that the dual vortex method at $ d=2 $ can achieve
    many important results which are very difficult to achieve from the direct boson picture.
    Although square lattice has been studied by various analytic and numerical methods before, the boson Hubbard model on
     a honeycomb lattice was not studied in any details by both analytic and
     Quantum Monte-Carlo (QMC) methods before the preprint \cite{univ}.
     It is not a Bravais lattice, so may show some different properties
     than those in square lattice.  Experimentally, the honeycomb lattice could also be easily realized in
     ultra-cold atomic experiments to be discussed in section II.
     The honeycomb lattice is also the relevant lattice for adatom adsorption on substrates
     to be discussed in section VIII.
     I found that a {\em supersolid  } ( SS ) state {\em exists only
     away from half filling}.
     A supersolid  in Eqn.\ref{boson} is a state with both
     superfluid and solid order. A supersolid in Eqn.\ref{spin} is defined as the
     simultaneous orderings of ferromagnet in the $ XY $ component ( namely, $ < b_{i} > \neq 0 $ )
     and CDW in the $ Z $ component.
     Recently, by using the torsional oscillator measurement, a PSU's group lead by Chan
     observed a marked $ 1 \sim 2 \% $ superfluid component
     even in bulk solid $^{4} He $ at $ \sim 0.2 K $  \cite{chan}.
     If this experimental observation indicates the existence of $^{4} He $ supersolid
     remains controversial \cite{qgl}. However,
     it was established by spin wave expansion \cite{gan}
     and quantum Monte-carlo (QMC) \cite{hard,add,soft} simulations that
     a supersolid state could exist in an extended boson Hubbard model (EBHM) with suitable lattice structures,
     filling factors, interaction ranges and strengths.

    It was also explicitly pointed out in \cite{univ} that the DVM developed in \cite{pq1}
    holds only in the superfluid ( SF ) and
     the valence bond solid ( VBS )  side where the saddle point of the dual gauge field can be taken as uniform,
     however, it fails in the charge density wave  ( CDW )
     side where the saddle point of the dual gauge field can {\em not} be taken as
     uniform anymore. So not only the fluctuations, but also the average values of the dual gauge
     fields are different on both sides.
     This is in sharp contrast to the superfluid to
     the Mott transition at and near $ f=n $ described by
     Eqns.\ref{vortex1} and \ref{vortex2} in the appendix.
     As explained below Eqn.\ref{vortex1}, the average density of bosons is the
    same in both the SF and the Mott insulating side, namely, it
    takes the integer $ n $ on both sides, so the average strength of the dual magnetic field can be
    taken as a uniform zero on both sides, of course,
    the fluctuations of the gauge field are completely different on both sides.
    However, in the CDW side which
    breaks the translational symmetry, special care is needed to choose
    a correct saddle point of the dual gauge field  in the CDW side to make the theory self-consistent,
    so a different action is needed in the CDW side \cite{univ}.
    This paper is an expanded version of the unpublished preprint \cite{univ}.
    In this expanded version, (1) by pushing the DVM
    to slightly away from commensurate filling factors and (2)
    also extending the DVM explicitly to the lattice symmetry breaking CDW side
    by choosing the corresponding  self-consistent saddle points of the dual gauge field,
    I will map out the global phase diagram of the EBHM on bipartite optical
    lattices such as honeycome and square lattice at and near half filling
    and also investigate superfluid, solid, especially supersolids and quantum phase transitions
    in the phase diagram in a {\sl unified } scheme.

        The DVM is a magnetic space group ( MSG ) \cite{pq1} symmetry-based approach
        which, in principle, can be used to classify all the possible phases
        and phase transitions after choosing correct saddle points for the dual gauge fields.
        But the question if a particular phase will appear or
        not as a ground state depends on the specific values of all the
        possible parameters in the EBHM in Eqn.\ref{boson}, so it
        can only be addressed by a microscopic approach such as Quantum Monte-Carlo
        (QMC) simulations. The DVM can guide the QMC to search for particular phases
        and phase transitions in a specific model. Finite size
        scalings in QMC in a  specific microscopic model can be used to confirm the phases and the universality
        classes of phase transitions discovered by the DVM. The two methods are complementary to each other and both are
        needed to completely understand phases
        and phase transitions in Eqn.\ref{boson}.

        The rest of the paper is organized as following,
        In section II,  I will explicitly derive the generators of
        the magnetic space group ( also called projective space
        group ) for $ q=2 $ in a honeycomb lattice, construct the
        effective actions which are invariant under this MSG, also
        write down both the charge density wave and the valence bond
        order parameters to characterize the symmetry breaking sides
        in the insulating regimes. In section III and IV, I will
        describe the phases and phase transitions driven by the competition of
        kinetic energy and repulsive potential energy in Fig. 4
        ( represented by the parameter $ r $ in Fig. 2 and Fig.3 ) along the
        horizontal axis at the commensurate fillings  and
        the phases and phase transitions driven by the chemical potential $ \mu $  along the
        vertical axis slightly away from the commensurate fillings
        in Ising and easy-plane limit respectively. We also stress
        that satisfying the self-consist condition for the dual gauge field in the
        Ising limit is absolutely necessary to achieve correct
        answers in the CDW side. In section V, we study the EBHM in a square
        lattice and also discuss the still disputed so called deconfined quantum critical point in the easy-plane limit.
        In section VI, we compare our results achieved by the DVM with some available QMC
        results and make implications on possible future QMC simulations in both lattices.
        Then we study the application of our
        results on 3 different experimental systems: ultra-cold atoms with long range interactions on
        optical lattices in both square and honeycomb lattice in section VII,
        adatom adsorption on different substrates
        such as  possible reentrant supersolids in the second layer of  $ ^{4}He $ adsorbed on
             graphite and Hydrogen adsorbed on Krypton-preplated
             graphite in honeycomb lattice in section VIII,
             possible cooper pair supersolid in high temperature superconductor
             $ La_{2-x} Ba_{x} Cu O_{4} $ in square lattice in section IX.
             finally, we reach conclusions in section X. In the
             appendix, we review the boson-vortex duality at integer
             filling $ f=n $ in which I will stress the role of the dual gauge field and
             its connection to the more general case of $ f=p/q $
             developed in \cite{pq1,univ} and the main text of this
             paper.




\section{ Magnetic space group,  effective action and order parameters in the dual vortex
picture.}

   In this section, we will extend the DVM in \cite{pq1} to
   study the EBHM Eqn.\ref{boson} in honeycomb lattice at and {\em slightly away } from $ q=2 $.
   Honeycomb lattice ( solid line in Fig.1 ) is not a Bravais lattice, so may show some different properties
   than a square lattice. The dual lattice of the honeycomb lattice is a triangular
   lattice ( dashed line in Fig.1 ).
   Two basis vectors of a primitive unit cell  of the triangular lattice can be chosen as
   $ \vec{a}_{1}=  \hat{x}, \vec{a}_{2}= - \frac{1}{2} \hat{x} + \frac{ \sqrt{3} }{ 2 } \hat{y},
    \vec{a}_{d}= \vec{a}_{1}+ \vec{a}_{2} $ as shown in Fig.1. The reciprocal lattice of a triangular
   is also a triangular lattice and spanned by two basis vectors
   $ \vec{k}= k_{1} \vec{b}_{1} + k_{2} \vec{b}_{2} $ with $ \vec{b}_{1}= \hat{x} + \frac{ \hat{y} }{ \sqrt{3} },
     \vec{b}_{2}= \frac{2}{\sqrt{3}} \hat{y} $ satisfying $ \vec{b}_{i} \cdot \vec{a}_{j} = \delta_{ij}  $.
    The point group of a triangular lattice is $ C_{6v} \sim D_6 $ which contains $ 12 $ elements. The two generators
    can be chosen as $ C_{6} = R_{\pi/3},  I_{1} $. The space group also includes the two translation operators
    $ T_{1} $ and $ T_{2} $ along $ \vec{a}_{1} $ and $ \vec{a}_{2} $ directions respectively.
    The 3 translation operators  $ T_{1}, T_{2}, T_{d} $, the rotation operator $ R_{\pi/3} $,
    the 3 reflection operators $ I_{1}, I_{2}, I_{d} $, the two rotation operators around the direct
    lattice points $ A $ and $ B $: $   R^{A}_{2\pi/3}, R^{B}_{2\pi/3} $ of the MSG
    are worked out in the following.

    In the Landau gauge $ \vec{A}=(0, Hx) $, the  mean field
    Hamiltonian for the vortices hopping in a triangular lattice
    in the presence of $ f $ flux quanta per triangle in the tight-binding limit is \cite{pq1} :
\begin{eqnarray}
  H_{0v}   & = &  -t \sum_{\vec{x}}  [  | \vec{x}+ \vec{a}_{1} > <  \vec{x} | + h.c.  \nonumber  \\
      & + & | \vec{x}+ \vec{a}_{2} > e^{i 2 \pi 2 f a_{1} } <  \vec{x} | + h.c.   \nonumber  \\
      & +  & | \vec{x}+ \vec{a}_{d} > e^{i 2 \pi 2 f ( a_{1} + 1/2 ) } <  \vec{x} | + h.c. ]
\label{hh}
\end{eqnarray}
   where $ \vec{x} = a_{1} \vec{a}_{1} + a_{2} \vec{a}_{2}   $ denotes lattice points of the triangular lattice.
   Note that the total vortex Hamiltonian $ H_{v} =H_{0v} + V $ where $ V $ is the interaction
   between vortices. Because $ V $ does not contain any
   Aharonov-Bohm (AB) phase factor from  the non-trivial $ A_{\mu} $ background, so the magnetic space group is
   completely determined by the vortex kinetic term $ H_{0v} $.  $ V $ always commutes with the generators
   of the MSG given by Eqns. \ref{space},\ref{reflect},\ref{direct}. So when constructing the representation
   of the MSG, we can ignore the $ V $ term without losing any generality.

    The  $ T_{1}, T_{2}, T_{d} $ and the rotation operator $ R_{\pi/3} $ of the PSG are worked out in \cite{pq1}.
   Here we listed them in slightly different notations:
\begin{eqnarray}
   T_{1}   & =  & \sum_{\vec{x}}| \vec{x}+ \vec{a}_{1} > e^{i 2 \pi 2f a_{2} } <  \vec{x} |  \nonumber  \\
   T_{2}   & =  & \sum_{\vec{x}}| \vec{x}+ \vec{a}_{2} > <  \vec{x} |    \nonumber  \\
   T_{d}   & =  & \sum_{\vec{x}}| \vec{x}+ \vec{a}_{d} > e^{i 2 \pi 2f ( a_{2}+1/2 ) } <  \vec{x} |  \nonumber  \\
    R_{\pi/3} & = & \sum_{\vec{x}}  e^{i 2 \pi f ( a^{2}_{1}- 2 a_{1} a_{2} ) } | a_{1}- a_{2} , a_{1} >< a_{1}, a_{2}|
\label{space}
\end{eqnarray}
     It can be shown that they all commute with $ H $. However, they do not commute with each other
     $ T_{1} T_{2}= \omega T_{d}, T_{1} T_{2}= \omega^{2} T_{2}T_{1} $.

   After performing some algebras, we found three of the 6 reflection operators in the point group $ C_{6v} $ :
\begin{eqnarray}
    I_{1} & = & K \sum_{\vec{x}}  e^{-i 2 \pi f a^{2}_{2} } | a_{1}- a_{2} , -a_{2} >< a_{1}, a_{2}|  \nonumber  \\
    I_{2} & = & K \sum_{\vec{x}}  e^{-i 2 \pi f a^{2}_{1} } | -a_{1}, a_{2}-a_{1} >< a_{1}, a_{2}|  \nonumber  \\
    I_{d} & = & K \sum_{\vec{x}}  e^{-i 2 \pi f 2 a_{1} a_{2} } | a_{2}, a_{1} >< a_{1}, a_{2}|
\label{reflect}
\end{eqnarray}
    where $ K $ is the complex conjugate operator which make $ I_{1}, I_{2}, I_{3} $  non-unitary operators.
    Note that in contrast to the reflection
    operators in square lattice, the phase factors in Eqn.\ref{reflect} are crucial to ensure they commute with
    the Hamiltonian Eqn.\ref{hh}.

    The eigenvalue equation $ H \psi (\vec{k} ) = E ( \vec{k} ) \psi (\vec{k} ) $ leads to the Harper's equation
   in the triangle lattice :
\begin{eqnarray}
    ( e^{-i k_{1} } +  e^{-i ( k_{1} + k_{2} + 2 \pi f ( 2l-1) ) } ) \psi_{l-1}( k_{1}, k_{2} )
                                      \nonumber  \\
      + ( e^{i k_{1} } + e^{i ( k_{1} + k_{2} + 2 \pi f ( 2l-1) ) } )
       \psi_{l+1}( k_{1}, k_{2} )   \nonumber  \\
    + 2 \cos(-k_{2} + 2 \pi f l ) \psi_{l}( k_{1}, k_{2} ) = E ( \vec{k} ) \psi_{l}( k_{1}, k_{2} )
\end{eqnarray}
    where $ l= 0,1,\cdots,q-1 $.

    In the following, we focus on $ q=2 $ case where there is only {\em one } band
   $ E( \vec{k} ) = - 2 t ( \cos k_{1} + \cos k_{2} - \cos ( k_{1} + k_{2} ) ) $. Obviously,
   $ E( k_1, k_2 ) = E( - k_1, -k_2 ) = E( k_2, k_1 ) $.  There are {\em two }  minima at $ \vec{k}_{\pm}=
   \pm ( \pi/3,\pi/3 ) $. Let's label the two eigenmodes at the two minima as
   $ \psi_{\pm} $.  By using the expressions of the rotation and reflection operators
   in Eqns. \ref{space}, \ref{reflect}, we find the two fields transform as:
\begin{eqnarray}
    T_{1}, T_{2} & : & \psi_{\pm} \rightarrow e^{\mp i \pi/3} \psi_{\pm};~~~
    T_{d}: \psi_{\pm} \rightarrow -e^{\mp i 2 \pi/3} \psi_{\pm}   \nonumber   \\
    R_{\pi/3} & : &  \psi_{\pm} \rightarrow  \psi_{\mp};~~~ I_{\alpha}: \psi_{\pm} \rightarrow \psi^{*}_{\mp},~~ \alpha=1,2,d
\label{trandual}
\end{eqnarray}
    where the transformations under $ T_{\alpha}, R_{\pi/3} $ were already derived in \cite{pq1}.
    Note that $ R_{\pi/3} $ plays the same role as the $ Z_2 $
    exchange symmetry between  $ \psi_{+} $ and $ \psi_{-} $.

    The quadratic terms of the effective action is simply the scalar electrodynamics as in the square lattice case:
\begin{equation}
    {\cal L}_{0} = \sum_{\alpha=\pm} | (  \partial_{\mu} - i A_{\mu} ) \psi_{\alpha} |^{2} + r | \psi_{\alpha} |^{2}
    + \frac{1}{4} F^{2}_{\mu \nu}
\label{scalar}
\end{equation}
   where $ F_{\mu \nu}=\frac{1}{2} \epsilon_{\mu \nu \lambda} \partial_{\nu} A_{\lambda} $ is the strength  of the non-compact gauge field $ A_{\mu} $.

   It is easy to show that there are only 2 independent quartic invariants under the above transformations:
\begin{equation}
    \Lambda_{1}  =  | \psi_{+} |^{4} + |\psi_{-} |^{4},~~~~ \Lambda_{2}  =  | \psi_{+} |^{2} |\psi_{-} |^{2}
\label{inv}
\end{equation}

    The direct honeycomb lattice is a non-Bravais lattice
    which contains two lattice points A and B per direct unit cell, it is useful to work out
    the rotation operators around the direct lattice points $ A $ and $ B $
    which contain new symmetries not included in the rotation operator around a dual triangular lattice
    point $ R_{\pi/3} $ :
\begin{eqnarray}
    R^{A}_{2\pi/3} & = & \sum_{\vec{x}}  e^{i 2 \pi f ( a^{2}_{2}- 2 a_{1} a_{2} - 2 a_{1} -2 a_{2} ) }
                             \nonumber  \\
    &   &  | - a_{2} + 1 , a_{1} - a_{2} >< a_{1}, a_{2}|  \nonumber  \\
    R^{B}_{2\pi/3} & = & \sum_{\vec{x}}  e^{i 2 \pi f ( (a_{2} - 1 )^{2}- 2 a_{1} a_{2} + 2 a_{1} ) }
                                                   \nonumber  \\
    &  &  | - a_{2} + 1 , a_{1} - a_{2} + 1 >< a_{1}, a_{2}|
\label{direct}
\end{eqnarray}

    They act on the two vortex fields $ \psi_{\pm} $ as:
\begin{equation}
    R^{A}_{2\pi/3}:  \psi_{\pm} \rightarrow  e^{\mp i \pi/3} \psi_{\pm};~~~~
    R^{B}_{2\pi/3}:  \psi_{\pm} \rightarrow  e^{\pm i \pi/3} \psi_{\pm}
\label{dir}
\end{equation}

   It is ease to see the two operators $ \Lambda_{1}, \Lambda_{2} $ in Eqn.\ref{inv} are also
   invariant under Eqn.\ref{dir}.
   Finally we reach the most general quartic term invariant under all the above transformations in Eqns.\ref{trandual},
   \ref{dir}:
\begin{equation}
    {\cal L}_{4}  =  \gamma_{0} ( | \psi_{+} |^{2} + |\psi_{-} |^{2} )^{2} -
                       \gamma_{1} ( | \psi_{+} |^{2} - |\psi_{-} |^{2} )^{2}
\label{inv4}
\end{equation}

      In the square lattice, In Landau gauge, a unitary transformation to the permutative representation is needed to reach
   Eqn.\ref{inv4}. Here in the gauge chosen in Eqn.\ref{hh}, $ \psi_{\pm} $ are automatically in the permutative
   representation. An important and subtle point is how to
   construct boson density order parameters to characterize the symmetry breaking patterns in the direct lattice
   in terms of the dual vortex fields in the dual lattice. In \cite{pq1}, the density order parameter
   was constructed to be the most general gauge invariant and {\em bilinear} combinations of the vortex fields
   $ \psi_{l} $. Then it was evaluated at the direct lattice points, links and dual lattice points
   to represent boson density, kinetic energy and amplitude of the ring exchanges respectively.
   The dual lattice of a square lattice is still a square lattice with the same lattice constant.
   The link points of a square lattice also form a square lattice with lattice constant $ \sqrt{2} $.
   Putting the direct, dual and link lattices together forms a square lattice with lattice constant $ 1/2 $.
   Although like a square lattice, the honeycomb lattice is also a bi-partisan lattice consisting of two interpenetrating
   triangular sublattices  $ A $ and $ B $, its dual lattice is a triangular lattice which is a frustrated one ( Fig.1),
   its link points form a Kagome lattice. Putting the three lattices together forms a very complicated lattice.
   So the density wave order parameters in the honeycomb lattice may not be evaluated in the three lattices.
   Indeed, when trying to identify the density operators in the insulating state,
   we find the density operators proposed in \cite{pq1} does not apply anymore in the honeycomb lattice.
   In the following, by studying how gauge invariant bilinear vortex fields transform under the complete PSG, we can identify
   both the boson density and boson kinetic energy operators in the direct lattice.
   We need evaluate these quantities {\em only} at dual lattice points to characterize the CDW and VBS
   orders in the direct lattice.
   Let's look at the {\em generalized } density order operators which characterize the symmetry breaking patterns in
   the insulating states. In the low energy limit, the vortex field
   is:
\begin{equation}
 \psi( \vec{x} )= e^{-i \vec{k}_{+} \cdot \vec{x} } \psi_{+}( \vec{x} ) +
   e^{-i \vec{k}_{-} \cdot \vec{x} } \psi_{-}( \vec{x} )
\end{equation}

   Intuitively, the generalized density operator
   $ \rho( \vec{x} )= \psi^{\dagger}( \vec{x} ) \psi( \vec{x} ) $ can be written as:
\begin{equation}
   \rho( \vec{x} ) = \psi^{\dagger}_{+} \psi_{+}
       + \psi^{\dagger}_{-} \psi_{-}
       + e^{i \vec{Q} \cdot \vec{x} }  \psi^{\dagger}_{+} \psi_{-}
       + e^{-i \vec{Q} \cdot \vec{x} }  \psi^{\dagger}_{-} \psi_{+}
\label{gen}
\end{equation}
    where $ \vec{Q}= 2 \pi/3 (1, 1) $.

   When taking continuum limit on the dual lattice, we take one site per dual unit cell which contains
   two direct lattice sites $ A $ and $ B $ ( Fig. 1).
   So Eqn.\ref{gen} should contain both the information
   on the boson densities on sites A and B and the boson kinetic energy on the link between A and B
   ( or $ XY $ exchange energy $ < S^{+}_{A} S^{-}_{B} + h.c. > $ in the spin language ).
   The single boson Green function hopping on the direct lattice is related to the gauge-invariant single vortex
   Green function on the dual lattice in a highly non-local way \cite{inv}. Fortunately, the two boson quantities such
   as the density and the kinetic energy may have simple local expressions in terms of the dual vortex fields.
   By studying how the operators in Eqn. \ref{gen} transform under \ref{trandual} and \ref{dir}, in the scaling limit,
   we can identify these quantities as ( up to a unknown prefactor) \cite{kondo} :
\begin{eqnarray}
   \rho_{A} & =  & \psi^{\dagger}_{+} \psi_{+},~~~ \rho_{B}=  \psi^{\dagger}_{-} \psi_{-}   \nonumber  \\
    K_{AB} & =  & e^{i \vec{Q} \cdot \vec{x} }  \psi^{\dagger}_{+} \psi_{-}
       + e^{-i \vec{Q} \cdot \vec{x} }  \psi^{\dagger}_{-} \psi_{+}
\label{iden}
\end{eqnarray}
    where $ \vec{x} $ stands for dual lattice points {\em only}.

\begin{figure}
\includegraphics[width=3cm]{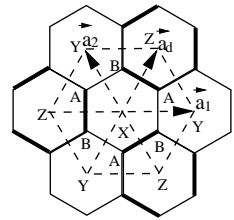}
\caption{ (a) Bosons at filling factor $ f $ are hopping on a
honeycomb lattice ( solid line ) which has two sublattices $ A $ and
$ B $. Vortices are hopping on its
    dual lattice which is a triangular lattice ( dashed line ) which has three sublattices $ X, Y, Z
    $. In the easy-plane limit shown in Fig.2c,
    one of the three VBS states \cite{vbs} on the honeycomb lattice is shown by the thick bonds in the figure.
   The other two VBS can be obtained by $ R^{A}_{ 2 \pi /3} $ or
    $ R^{B}_{ 2 \pi /3} $. }
\label{fig1}
\end{figure}

    Moving {\em slightly} away from half filling $ f=1/2 $ corresponds to adding
    a small {\em mean} dual magnetic field $ H \sim  \delta f= f-1/2 $ in the action.
    It can be shown that {\em inside the SF phase}, the most general
    action invariant under all the MSG transformations  upto quartic terms is :
\begin{eqnarray}
    {\cal L}_{SF} & = &
    \sum_{\alpha=a/b} | (  \partial_{\mu} - i A_{\mu} ) \psi_{\alpha} |^{2} + r | \psi_{\alpha}
    |^{2}     \nonumber   \\
   & + & \frac{1}{4 e^{2} } ( \epsilon_{\mu \nu \lambda} \partial_{\nu} A_{\lambda}
    - 2 \pi \delta f \delta_{\mu \tau})^{2}
    + {\cal L}_{4}
\label{away}
\end{eqnarray}
     where $ A_{\mu} $ is a non-compact  $ U(1) $ gauge field.
     Upto the quartic level, with correspondingly defined $ \psi_{a/b} $ in a square lattice,
     Eqn.\ref{away} is {\sl the same } as that in the square lattice derived in \cite{pq1}.
     Because Eqn.\ref{away} is a long wavelength effective action, the relations between the phenomenological
     parameters in Eqn.\ref{away} and the microscopic parameters in
     Eqn.\ref{boson} are not known.
     Fortunately, we are still able to classify some
     phases and phase transitions and make some concrete predictions from Eqn.\ref{away} without knowing
     these relations.
    In the following, we assume $ r < 0 $ in Eqn.\ref{scalar}, so the system is in insulating states.
    We  will discuss the Ising limit and the easy-plane limit respectively.

\section{Ising limit }

   If $ \gamma_{1} > 0 $, the system is in the
   Ising limit, the mean field solution is $ \psi_{a} =1, \psi_{b}=0
   $ or vice versa. The system is in the  CDW  order.
   We can see how $ \rho_{A}, \rho_{B} $ transform under Eqns. \ref{trandual} and \ref{dir}:
\begin{equation}
   T_{\alpha}, R^{A/B}_{2 \pi/3}: \rho_{A/B} \rightarrow \rho_{A/B};~~~
   R_{\pi/3}, I_{\alpha}:  \rho_{A} \leftrightarrow \rho_{B}
\label{den}
\end{equation}

   These transformations confirm that $ \rho_{A} $ and $ \rho_{B} $ indeed can be identified as the boson density
   operators at direct sublattices $ A $ and $ B $. In this section, we discuss at and away from half filling
   respectively.

\subsection{  SF to CDW transition at half filling $ \delta f=0 $. }

   If $ r < 0 $,  the system is in the CDW order
   which could take checkboard $ (\pi,\pi) $ order  or a stripe order \cite{equal}.
   Eqn.\ref{away} is an expansion around the  uniform saddle point $
   < \nabla \times \vec{A} > = f = 1/2 $ which holds in the SF and the
   VBS ( to be discussed in section 4 ), so it breaks down in the CDW side.
   So it can not be used to study the SF to the CDW transition.

   In the CDW state where $ r < 0 $, a
   different saddle point where  $ <\nabla \times \vec{A}^{a}> = 1-\alpha
   $ for sublattice $ A $ and $ < \nabla \times \vec{A}^{b}> = \alpha
   $ for sublattice $ B $  with $ \alpha < 1/2 $ should be used.
   It is easy to see  that there
   is only one vortex minimum $ \psi_{b} $ in the staggered dual magnetic
   field at $ \delta f=0 $, the effective action {\em inside} the CDW state is:
\begin{eqnarray}
  {\cal L}_{C-CDW} & = & | (  \partial_{\mu} - i A^{b}_{\mu} ) \psi_{b} |^{2} + \tilde{ r } |
  \psi_{b}|^{2} +  \tilde{u}  | \psi_{b} |^{4} + \cdots     \nonumber  \\
     & + & \frac{1}{ 4 \tilde{e}^{2} } ( \epsilon_{\mu \nu \lambda} \partial_{\nu} A^{b}_{\lambda} )^{2}
\label{is0}
\end{eqnarray}
   where $ \tilde{r} < 0 $, so the system is in the CDW state where $ < \psi_b > \neq 0
   $. Eqn.\ref{is0} is essentially the same as Eqn.\ref{vortex1} in
   $ q=1 $ case.
   Note that because $ < \psi_a > \neq 0 $, the gauge field $ \vec{A}^{a} $  is always massive, so
   it does not appear in Eqn.\ref{is0}. As explained in the
   appendix, this indicates the density fluctuations in sublattice $ A
   $ is suppressed, so can be taken as fixed.

   Due to the change of the saddle point structures,
   the transition from the SF to the CDW driven by the horizontal axis ( quantum fluctuation $ r $ ) in Fig.2b
   is a {\em strong} first order transition.
   If we assumed that Eqn.\ref{away} at $ \delta f=0 $
   in the Ising limit could be used to describe this first order
   transition, then the CDW side could be described by $ < \psi_a > \neq 0, < \psi_b > = 0 $
   or vice versa \cite{kagome}, but no phases with  $ < \psi_{a} > \neq 0, < \psi_b > \neq 0  $.
   However, as shown in Eqn.\ref{is0}, in the CDW side, $
   < \psi_{a} > \neq 0, < \psi_b > \neq 0 $, so both gauge fields $ \vec{A}^{a} $
   and $ \vec{A}^{b} $  are massive, the density fluctuations in both sublattice $ A
   $  and $ B $ are suppressed, so both  can be taken as fixed.  This fact is due to the
   change of saddle point structure across the SF to the CDW
   transition. So we conclude that due to this change of saddle point structure of the dual gauge
   fields across the SF to the CDW, Eqn.\ref{away} at $ \delta f=0 $
   in the Ising limit can not be used to describe this first order transition.
   However, it does give a  qualitative indication of this strong
   first order transition. This is expected, because in a strong first
   order transition, two separate dual actions Eqn.\ref{away} in the SF side and Eqn.\ref{is0}
   are needed to describe the two sides separately.
   Note that in the direct picture, the SF breaks the $ U(1) $
   symmetry, the CDW breaks the lattice symmetry, so the two sides
   break two completely different symmetries, it can only be first
   order.

\subsection{ Charge Density Wave (CDW) supersolid away from half
filling $ \delta f \neq 0 $.}

   Now we look at the effects of the in-commensurability $ \delta f=
   f-1/2 $ in Eqn.\ref{is0}. In the SF side where $ r > 0 $, it
   is known the SF is stable against the change of the chemical
   potential ( or adding bosons ). In the CDW side where $ \tilde{r} < 0 $,
   moving {\em slightly} away from half filling $ f=1/2 $ corresponds to adding
   a small {\em mean} dual magnetic field $ H \sim  \delta f= f-1/2 $ in
   Eqn.\ref{is0}:
\begin{eqnarray}
  {\cal L}_{IC-CDW} & = & | (  \partial_{\mu} - i A^{b}_{\mu} ) \psi_{b} |^{2} + \tilde{ r } |
  \psi_{b}|^{2} +  \tilde{u}  | \psi_{b} |^{4} + \cdots     \nonumber  \\
     & + & \frac{1}{ 4 \tilde{e}^{2} } ( \epsilon_{\mu \nu \lambda} \partial_{\nu} A^{b}_{\lambda}
    - 2 \pi \delta f \delta_{\mu \tau})^{2}
\label{is}
\end{eqnarray}
   where the vortices in the phase winding of $ \psi_{b} $  should be interpreted as the
   the boson number \cite{direct}. Note that because
   $ < \psi_a > \neq 0 $ and remains non-zero  in the presence of $ \delta f $,
   the gauge field $ \vec{A}^{a} $  is always massive and remains massive in the presence of $ \delta f $, so
   it still does not appear in Eqn.\ref{is0}. This indicates the density fluctuations in sublattice $ A
   $ remains suppressed, so can be taken as fixed in the presence of $ \delta f $.

   Eqn.\ref{is} is essentially the same as Eqn.\ref{vortex2} which
   has the structure identical to the conventional $ q=1 $ component
   Ginzburg-Landau model for type-II " superconductors "  in a "magnetic" field.
   It was shown in \cite{huse} that
   for type II superconductors, the gauge field fluctuations will render the vortex fluid phase
   intruding at $ H_{c1} $ between the Messiner and the mixed phase ( see Fig. 2a).
   For parameters appropriate to the cuprate
   superconductors, this intrusion occurs over too narrow an interval of $ H $ to be observed in experiments.
   In the present boson problem with the nearest neighbor interaction $
   V_{1} > 0 $ in Eqn.\ref{boson} which stabilizes
   the $ (\pi,\pi) $ CDW state at $ f=1/2$, along the dashed line
   driven by the vertical axis ( chemical potential $ \mu $ ) in
   Fig.2b, this corresponds to a CDW  supersolid (CDW-SS) state intruding between
   the commensurate CDW state at $ f=1/2 $ and
   the in-commensurate CDW state at $ 1/2 + \delta f $ which could be stabilized by further neighbor
   interactions in Eqn.\ref{boson}.
   The first transition is in the $ z=2, \nu=1/2, \eta=0 $ universality
   class first discussed in \cite{boson}, while the second is a 1st order transition.
   We expect the intruding window at our $ q=2 $ system is much  wider  than that of the $ q=1 $ system.
   In the CDW states, $ < \psi_{b} > \neq 0 $, so the dual gauge
   field $ \vec{A}^{b}_{\mu} $ in Eqn.\ref{is} is massive.
   In the CDW-SS state, $ < \psi_{b} > =0 $, there is the
   gapless superfluid mode represented by the dual gauge field $
   A^{b}_{\mu} $. The CDW-SS has the same $ ( \pi, \pi ) $ diagonal order
   as the C-CDW.
   The identified universality class of the CDW to the CDW-SS transition has
   many physical implications. For example, near the C-CDW to the CDW-SS transition, the superfluid
   density  should scale as $ \rho_{s} \sim |\rho-1/2|^{(d+z-2)\nu }=|
   \rho-1/2|= \delta f $ with a logarithmic correction.
   The logarithmic correction will be calculated in a future publication .
   There must be a transition from the CDW-SS to the SF inside the window driven by
   the quantum fluctuation $ r $ in the Fig.2b. The universality
   class of this transition is likely to be first order and will be investigated further in a future
   publication . Combining the results in III-A and III-B leads to the
   global phase diagram Fig.2b in the Ising limit \cite{bi}.

\begin{figure}
\includegraphics[width=8cm]{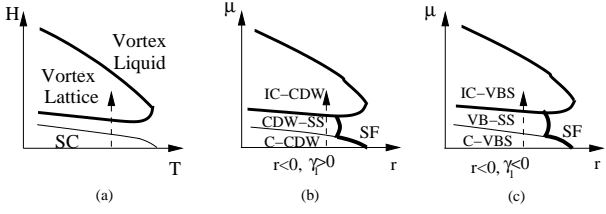}
\caption{(a) The phase diagram {\em slightly away}
    from $ f=1 $  described by Eqn.\ref{vortex2} in the appendix is the same as type-II superconductors in an external magnetic field $ H $  where there
    should be a vortex liquid state intruding between the Messiner state and the
    vortex lattice state. But the intruding regime is too narrow to be seen in type-II
    superconductors \cite{huse}.  (b) and (c) are the  zero temperature global phase diagrams of
   the chemical potential $ \mu $ versus $ r $ in Eqn.\ref{away} in the honeycomb lattice.
   (b) The Ising limit $ \gamma_{1} > 0 $ discussed in section III. There is a CDW  supersolid
   (CDW-SS) state intruding between the commensurate
   CDW ( C-CDW ) state at $ f=1/2 $ and the in-commensurate CDW (IC-CDW) state at $ 1/2 + \delta f $.
    The CDW-SS
    has the same lattice symmetry breaking as the C-CDW.
   (c) The Easy-Plane limit $ \gamma_{1} < 0 $ discussed in section IV.
   There is a Valence Bond Supersolid (VB-SS) state intruding between the commensurate
   VBS (C-VBS) state at $ f=1/2 $ and the in-commensurate VBS ( IC-VBS ) state at $ 1/2 +
   \delta f $. The VB-SS  has the same lattice
   symmetry breaking as the C-VBS. The thin ( thick ) line is the 2nd ( 1st ) order transition.
   The 1st order transition in the Ising ( Easy-plane ) limit is {\em strong } ( {\em weak} )
   one. (b) and (c) are drawn only above half filling. $ \mu \rightarrow -
   \mu $ corresponds to below half filling. }
\label{fig2}
\end{figure}


\section{ Easy-plane limit.}

   If $ \gamma_{1} < 0 $, the system is in the easy-plane limit, the mean field
   solution is $ \psi_{a}=e^{i \theta_{a} }, \psi_{b}= e^{i \theta_{b}}
   $, then $ \rho_{A}= \rho_{B} = 1 $, so the two sublattices
   remain equivalent.  This limit could be reached by possible ring exchange
   interactions in Eqn.\ref{boson}. In the following, we discuss at and away from half filling
   respectively.

\subsection{  SF to VBS transition at half filling $ \delta f=0 $. }

   Because the two sublattices remain
   equivalent, the uniform saddle point $ < \nabla \times \vec{A} > = f = 1/2 $
   holds in both the SF and the VBS.
   The system has a VBS order with the  kinetic energy order parameter $ K_{AB} =  \cos( \vec{ Q } \cdot \vec{x} +
   \theta_{-} ) $ where $ \theta_{-} = \theta_{a} - \theta_{b} $.
   Let's look at how the kinetic energy $ K $ transform:
\begin{eqnarray}
   T_{\alpha} & : & K \rightarrow \cos( \vec{Q} \cdot (\vec{x} - \vec{a}_{\alpha}  )  + \theta )    \nonumber  \\
   R_{\pi/3}  & : & K \rightarrow \cos( \vec{Q} \cdot \vec{x} - \theta ); ~~~ I_{\alpha}:  K \rightarrow  K  \nonumber  \\
   R^{A/B}_{ 2 \pi/3} & : &  K \rightarrow \cos( \vec{Q} \cdot \vec{x} + \theta \mp 2 \pi/3 )
\label{kin}
\end{eqnarray}
   Note that $ K $ transforms differently under $ R_{\pi/3} $ and $ I_{\alpha} $, because the latter are anti-unitary
   operators.  These transformations confirm that $ K $ indeed can be identified as the boson
   kinetic energy or the $ XY $ exchange energy operators at the direct lattice.

   Upto the quartic order,
   the relative phase between $ \psi_{+} $ and $ \psi_{-} $ is undetermined. Higher order terms are needed
   to determine the relative phase.  It is clear to see there are only 3 sixth order invariants:
\begin{eqnarray}
    C_{1}   & =  & | \psi_{+} |^{6} + |\psi_{-} |^{6}   \nonumber  \\
    C_{2}   & =  &  ( | \psi_{+} |^{2} + |\psi_{-} |^{2} ) | \psi_{+} |^{2} |\psi_{-} |^{2}  \nonumber  \\
    C_{3}   & = &  \lambda [ ( \psi^{*}_{+} \psi_{-} )^{3}  + ( \psi^{*}_{-} \psi_{+} )^{3} ]
     = \lambda \cos 3 \theta
\label{six}
\end{eqnarray}
   where $ \theta= \theta_{+}-\theta_{-} $. Especially $ C_{3} $ is invariant under $ R^{A/B}_{ 2 \pi/3} $
   Obviously, only the last term $ C_{3} $  can fix the relative
   phase. This term corresponds to the 3-monopole operator in the spin language.

   If $ \lambda > 0, \theta= (2n+1) \pi/3= \pi/3, \pi, - \pi/3  $.
   The kinetic energy $ K= 1, -2, 1 $ takes only two values, one strong bond and two
   weak bonds, their ratio is 2.  The dual triangular lattice can be
   divided into three sublattices $ X, Y, Z $ where $ x_{1}+ x_{2} =0, 1, 2 $ mod $ 3 $, hence $ \vec{Q} \cdot \vec{x}=
   0, 2 \pi/3, -2 \pi/3 $ respectively ( Fig.1). If we evaluate
   $ K $ at $ X, Y, Z $, we find $ K_{X}= \cos \theta, K_{Y}= \cos ( \theta + 2 \pi/3), K_{Z}= \cos ( \theta - 2 \pi/3) $.
   If we know one bond in one dual unit cell, say, labeled by $ X $, then by the translation, we can get all the other bonds
   in sublattices $ Y $ and $ Z $ with the same orientation. The two bonds on the other two orientations can be reached by
   rotations listed in Eqn. \ref{kin}. By this way, we can get all the bonds
   in the whole direct lattice. So the system is in the Valence Bond Solid (VBS ) state, one VBS is shown in Fig.1.

   If $ \lambda < 0, \theta=  2n \pi/3 = 0, 2 \pi/3, - 2 \pi/3  $.
   The kinetic energy $ K= 2, -1, -1 $ takes also only two values, one strong bond and two
   weak bonds, their ratio is also 2. This case is essentially the same as $ \lambda > 0 $.
   Because the sign of the kinetic term can be changed in a bi-partisian lattice
    by changing the sign of $ b_{i} $ in one of the two sublattices in Eqn.\ref{boson}
    (or  can be changed by changing the sign of $ S^{x}_{i}, S^{y}_{i} $ in one of the two sublattices,
    but keeping $ S^{z}_{i} $ untouched in Eqn.\ref{spin} ),
   but the product of the sign around a hexagon is fixed. The two cases have the same sign product, so
   can be transformed to each other by the transformation. This can also be understood by observing that
   $ C_{3} $ in Eqn.\ref{six} behaves like a hopping term in 3 power, so its sign
   can be changed by the transformation. This is in sharp contrast to the square lattice to be discussed
   in section V where one sign leads to a
   columnar dimer, the other leads to a plaquette pattern.

\subsection{ Valence Bond Supersolid away from half filling $ \delta f
\neq 0 $.}

   Now we look at the effects of the in-commensurability $ \delta f=
   f-1/2 $ in Eqn.\ref{away}. Again, the SF side where $ r >0 $ is stable against the change of the chemical
   potential. In the VBS side where $ r < 0 $, slightly away from the half-filling, Eqn.\ref{away} becomes:
\begin{eqnarray}
  {\cal L}_{VBS}  &  =  & ( \frac{1}{2} \partial_{\mu} \theta_{+} - A_{\mu} )^{2}
      + \frac{1}{4 e^{2}} ( \epsilon_{\mu \nu \lambda} \partial_{\nu} A_{\lambda}
      - 2 \pi \delta f \delta_{\mu \tau})^{2} + \cdots    \nonumber  \\
       & + &  ( \frac{1}{2} \partial_{\mu} \theta_{-} )^{2} + 2 \lambda \cos 3 \theta_{-}
\label{ep1}
\end{eqnarray}
   where $ \theta_{\pm} = \theta_{a} \pm \theta_{b} $.

   Obviously, the $ \theta_{-} $ sector is massive ( namely, $ \theta_{a} $ and $ \theta_{b} $ are {\em locked} together )
   and can be integrated out. Assuming $ \lambda > 0 $, then $ \theta_{-} = \pi $.
   Setting $ \psi_{+}=e^{i \theta_{+} } \sim \psi_{a} \sim -\psi_{b}$ in Eqn.\ref{away} leads
   to Eqn.\ref{is} with $ \tilde{ u }  = 2 \gamma_{0} $, so
   the discussions on Ising limit case following Eqn.\ref{is} also
   apply. In the present boson problem with possible ring exchange interactions
   in Eqn.\ref{boson} which
   stabilizes the VBS state at $ f=1/2$, along the dashed line
   driven by the vertical axis ( chemical potential $ \mu $ ) in
   Fig.2c, this corresponds to a VBS supersolid (VB-SS) state intruding between the commensurate
   VBS ( C-VBS) state at $ f=1/2 $ and the in-commensurate VBS (IC-VBS) state at $ 1/2 +
   \delta f $ as shown in Fig.2c.
   In the C-VBS state described in the subsection A, $ < \psi_{a} > = - < \psi_{b} >
   \neq 0, < \psi^{\dagger}_{a} \psi_{a} > = < \psi^{\dagger}_{b} \psi_{b} > =
   - < \psi^{\dagger}_{a} \psi_{b} > \neq 0 $, the gauge field $
   A_{\mu} $ is massive due to the Higgs mechanism.
   In the VB-SS state, $ < \psi_{a} > = < \psi_{b} > =0 $,
   but $ < \psi^{\dagger}_{a} \psi_{a} > = < \psi^{\dagger}_{b} \psi_{b} > = - < \psi^{\dagger}_{a} \psi_{b} > \neq 0 $,
   so there is a VBS order characterized by the order parameter
   $ K_{AB} =  \cos( \vec{ Q } \cdot \vec{x} + \theta_{-} )
   $ which is the same as the C-VBS, while the gauge field $ A_{\mu} $ is
   massless which stands for the gapless superfluid mode inside the VB-SS.
   So this VB-SS has both the VBS order and the superfluid order
   which justifies its name. {\em So in this VB-SS, the density
   fluctuation on each site is very large signalizing its superfluid
   nature, while the VBS order is fixed signalizing its VBS
   nature }. In this IC-VBS state, $ \delta f $
   valence bonds shown in Fig.1 is slightly stronger than the
   others, these $ \delta f $ slightly stronger bonds also form a
   dilute lattice on top of the underlying C-VBS lattice.
   Again, the first transition  from the C-VBS to the VB-SS is in the  $ z=2, \nu=1/2, \eta=0 $ universality
   class, while the second  from the VB-SS to the IC-VBS is 1st order. Near the C-VBS to the VB-SS transition, the superfluid
   density  should scale as $ \rho_{s} \sim |\rho-1/2|^{(d+z-2)\nu }=|
   \rho-1/2|= \delta f $ with logarithmic corrections.
   The nature of the transition from the VB-SS to the SF where $ < \psi_{a} > = < \psi_{b} >
   =0 $ and  $ < \psi^{\dagger}_{a} \psi_{b} > = 0 $  inside the window driven by
   the quantum fluctuation $ r $ in the Fig.2c is likely to be weakly first order
   and will be studied  further in a future publication. Combining (4a) and
   (4b) leads to the global phase diagram Fig.2c.


\section{  Square lattice. }

     As said below Eqn.\ref{away}, with correspondingly defined $ \psi_{a/b} $ in a square lattice,
     upto the quartic level, Eqn.\ref{away} is the same as that in the
     square lattice derived in \cite{pq1}.

\subsection{Ising limit}

     So the phase diagram in the Ising limit Fig.3b
     remains the same as Fig.2b. The SF to the C-CDW transition  at $ \delta f = 0 $
     along the horizontal axis is a {\em strong} first order one.
     Away from the half filling, along the dashed line in Fig.3b,
     the IC-CDW  can be stabilized only by very long range
     interactions in Eqn.\ref{boson}, if it is not stable, then Fig.3b reduces to
     Fig. 4b ( which is the Fig.14 in \cite{add}).

\subsection{Easy-plane limit}

     In the Easy-plane limit, as shown in \cite{pq1}, the lowest order
     term coupling the two phases $ \theta_{a/b} $ is
     $  C_{sq}= \lambda \cos 4 \theta $,  so the $ C_{3} $ term in
     Eqn.\ref{ep1} need to be replaced by $ C_{sq} $.
     If $ \lambda $ is positive ( negative),
     the VBS is Columnar dimer ( plaquette ) pattern \cite{pq1}.

\begin{figure}
\includegraphics[width=8cm]{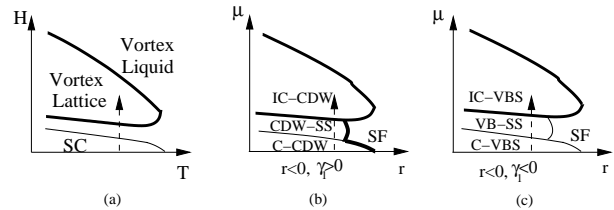}
\caption{ Phase diagram in square lattice. (a) is identical to 2a
and presented here just for completeness. (b) is similar to Fig.2b
on the honeycomb lattice in the Ising limit. In the easy plane limit
(c), there is a possible 2nd order transition between the SF and the
C-VBS through a so called deconfined quantum critical point.
However, this has been disputed in \cite{umass}. }. \label{fig3}
\end{figure}

   By using duality analysis and QMC simulation, the author in \cite{decon} suggested
   that in the easy plane limit, the $ q=2 $ component scalar electrodynamics Eqn.\ref{away}
   at $ \delta f=0 $ along the horizontal axis  is a second order transition through a
   so called "de-confined QCP". If this is indeed the case, then  $  C_{sq}= \lambda \cos 4 \theta $
   is irrelevant, there is a possible 2nd order transition between the SF and the C-VBS
   through a so called deconfined quantum critical point as shown in Fig.3c.
   Away from the half filling, along the dashed line in Fig.3c, the
   phase transitions are the same as those in the honeycomb lattice
   Fig. 2c. The transition from the VB-SS to the SF where $ < \psi_{a} > = < \psi_{b} >
   =0 $ and  $ < \psi^{\dagger}_{a} \psi_{b} > = 0 $  inside the window driven by
   the quantum fluctuation $ r $ in the Fig.3c is also through line
   of deconfined quantum critical points.  Recently, some connections are made between the correlation
  functions ( which lead to transport properties such as conductivity ) near the deconfined
  quantum critical point  in Fig.3c and those in  $ {\cal N}=8 $
  supersymmetric $ SU(N) $ Yang-Mills gauge theory in the large $
  N $ limit at $ d=2 $\cite{tran}. These correlation functions were calculated through
  $ AdS_{4}  \times S^{7}/CFT_{3} $ connection \cite{tran}. The particle-vortex
  self-duality  in Fig.3c and corresponding
  electro-magnetic self-duality in $ AdS_{4} \times S^{7}/CFT_{3} $
  put strong constraints on correlation functions in the two models respectively.

   However, recently, by using QMC simulations
   at much larger systems, the authors in \cite{umass} pointed out
   that the QMC study in \cite{decon} violates the hyperscaling and
   concluded that there is no such deconfined QCP, so the transition from SF to the VBS driven
   by the horizontal axis ( quantum fluctuation $ r $ ) is  a {\em weak} first order one.
   If this is indeed the case, the phase diagram in the easy-plane limit
   remains the same as in a honeycomb lattice Fig.2c.


\section{ Implication on Quantum Monte-Carlo (QMC) simulations.}

   The EBHM Eqn.\ref{boson} of the hard core bosons on square lattice with  $ V_{1} $ and
   $ V_{2} $ interactions was studied by the  QMC simulations in \cite{add},
   the authors found a stable striped $ ( \pi,0) $ and $ (0,\pi) $
   SS ( Fig.4 ). A stable $ ( \pi, \pi) $ SS can only be realized
   in the soft core boson case \cite{soft}.
   But the nature of the CDW to supersolid transition has never
   been addressed. Our results in the Fig.3 show that the CDW to the SS transition must be in
   the same universality class of Mott to superfluid transition with exact exponents $ z=2, \nu=1/2, \eta=0
   $ with a logarithmic correction. It is important to  (1) confirm this prediction by finite
   size scaling through the QMC simulations in square lattice for $ (0,\pi) $ and $ (\pi,0 ) $
   supersolid in both hard core and soft core  case in the Fig.4 and $ (\pi,\pi) $ supersolid in the soft core
   case only
   (2) do similar things in honeycomb lattice to confirm Fig.1. Some
   of our results in Fig.3 are indeed confirmed in very recent QMC simulations on soft core
   bosons in honeycomb lattice \cite{qmc}.
   (3) To Eqn.\ref{boson} with $ U=\infty, V_{1} > 0 $, adding a ring exchange term $ -K_{s} \sum_{ijkl} ( b^{\dagger}_{i}
   b_{j} b^{\dagger}_{k} b_{l} + h.c. ) $  where $ i, j, k, l $ label
   4 corners of a square in the square lattice and $ -K_{h} \sum_{ijklmn} ( b^{\dagger}_{i}
   b_{j} b^{\dagger}_{k} b_{l} b^{\dagger}_{m} b_{n} + h.c. ) $ where $ i, j, k, l, m, n $ label
   6 corners of a hexagon in the honeycomb lattice to stabilize the C-VBS state {\em  at half filling
   } (  Note that if $ K_{s}, K_{h} > 0 $, the QMC are free of sign problems),
   then confirm the prediction on C-VBS to VB-SS transition in Fig.2c and Fig.3c.
   We expect that  the
   VB-SS phase should be stable against a phase separation in both hard and soft core cases.
   The second transition ( CDW-SS to IC-CDW in Fig.2b and Fig.3b and
   the VB-SS to IC-VBS in Fig. 2c and Fig.3c ) is hard to be tested in QMC, because some very long
   range interactions are needed to stabilize the IC-CDW or the IC-VBS
   state. They are first order transition anyway.

\begin{figure}
\includegraphics[width=6cm]{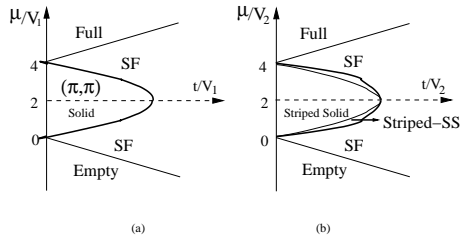}
\caption{
    Phase diagrams of the EBHM Eqn.\ref{boson} in the hard core limit $ U=\infty $
    achieved from the QMC in \cite{add}:
    (a) with the nearest neighbor interaction $ V_{1}  > 0 $, the
    CDW ordering wavevector is $ ( \pi, \pi ) $. The corresponding
    $ (\pi,\pi) $ supersolid is  unstable against the phase separation.
    However, it becomes stable in the soft core limit \cite{soft}
    (b) $ V_{1}=0 $, but next nearest neighbor interaction $ V_{2} > 0$,
    the CDW ordering wavevector is $ ( \pi, 0 ) $ or $ ( 0, \pi) $ which is called stripe phase.
    The corresponding stripe supersolid is stable even in the hard
    core limit. So there is a narrow window of stripe SS intervening
    between the stripe solid and the SF.
    The universality class of the stripe solid to the stripe SS  in (b)
    was not studied in \cite{add}. The thin (thick) line  is a 2nd (1st ) order transition. }
\label{fig4}
\end{figure}

   In fact, one of the predictions in this paper on the scaling of the superfluid
   density $ \rho_{s} \sim | \rho-1/2| $  was
   already found in the striped $ (\pi,0) $ solid to striped
   supersolid transition by QMC in Sec.V-B of Ref. \cite{add}. As shown in section III-B, there should
   be a logarithmic correction to the scaling of $ \rho_{s} $, it
   remains a challenge to detect the logarithmic correction by a high precision QMC.
   Of course, the superfluid density is anisotropic $ \rho^{x}_{s}> \rho^{y}_{s} $ in the $ (\pi,0) $
   supersolid, but they scale in the same way with different coefficients
   \cite{add}. Although the
   authors in \cite{add} suggested it is a 2nd order transition, they did not
   address the universality class of the transition.

   In the following three sections, I will discuss the applications
   of the results achieved in the previous sections on 3 different
   experimental systems.

\section{ Ultra-cold atom on square and honeycomb optical lattices }

      Recently, Bose-Einstein condensation
     (BEC) was realized in ultra-cold atomic gases ( for a review , see \cite{rev} ).
      Superfluid to Mott insulator transition was also observed in optical lattices of ultra-cold alkali  atoms \cite{cold}.
      Atomic physicists are constructing effective
      various kinds of 2d and 3d optical lattices using laser beams and then load either ultra-cold fermion
      or boson atoms at different filling factors on the lattices.
      They may tune the parameters in Eqn.\ref{boson} to realize different phases and quantum phase transitions
      \cite{cold,rev}. The optical honeycomb lattice geometry Fig.1 and square lattice could be realized in
      future ultra-cold atomic experiments. The challenge is to
      achieve longer range interactions than the onsite interaction.
     Very exciting perspectives to achieve longer range interactions have been opened by recent
     experiments \cite{polar} on cooling and trapping of polar molecules.
     Being electrically or magnetically polarized, polar molecules
     interact with each other via long-rang anisotropic
     dipole-dipole interactions. Loading the polar molecules on a 2d optical lattice
     \cite{polarlattice} with the dipole moments
     perpendicular to the trapping plane can be mapped to
     Eqn.\ref{boson} with long-range  repulsive interactions $ \sim p^{2}/r^{3}
     $ where $ p $ is the dipole moment. There are also other ways to generate long-range interactions \cite{qs}.
     We expect Fig.2a can be easily realized. There could also be some
     efficient ways to generate the ring exchange interactions \cite{qs}
     needed to stabilize the the VBS state in Fig.2b.

\section{ Adatom adsorptions on substrates ( honeycomb lattice )  }

  In this section, we will use the phase diagram Fig.2a to discuss two
  experimental systems: (1) the reentrant "superfluid" detected in a narrow region of coverages
  in the second layer of He4 ( called $^{4}He/^{4}He$/graphite
  system ) adsorbed on graphite by a previous torsional oscillator experiment (  \cite{he}.
  (2) the reentrant "fluidlike" state of hydrogen adsorbed on Krypton-preplated graphite ( $ H_2
  $/Kr/graphite ) near half filling which was investigated in a recent experiment \cite{honey}.

        The dual vortex approach is a Magnetic Space Group ( MSG ) symmetry based approach which can be used to classify all the possible phases
        and phase transitions. However, the question if a particular phase will appear or
        not as a ground state depends on the specific values of all the
        possible parameters in the boson Hubbard model in
        Eqn.\ref{boson}, namely, the specific values in the real systems,
        it can not be addressed in this
        approach. There are two ways to remedy this short-coming.
        (1) we can compare our theoretical classifications with some known phases observed in experiments, then
        we can be more specific on our predictions on the nature of {\em unknown} phases and phase transitions.
        (2) As said in section VI, a microscopic approach such as Quantum Monte-Carlo (QMC) may be
        needed to supplement the dual field theoretical approach .
        In this section, we will take the first
        strategy to study the boson Hubbard model Eqn.\ref{boson} in honeycomb
        lattice Fig.1 near $ q=2 $ ( Fig.1 ). The honeycomb lattice in Fig.1 may describe
        the preferred adsorption sites in the two systems.
        So the $^{4}He/^{4}He$/graphite and $ D_2 $/Kr/graphite systems may have
        the same symmetry and belong to the same universality class.

\subsection{ $^{4}He/^{4}He$/graphite system }

     A superfluid is a fluid that flows through the tiniest channels or cracks without viscosity.
     So far, the phenomenon of superfluidity has been firmly observed in only two kinds of systems.
     The first system is the two isotopes of
     Helium: $^{4}He $ and $^{3}He$. $ He4$ is a boson and becomes a superfluid when
     $ T < T_{c} =2.18 K $, while $ He3 $ is a fermion, two $ He3 $ atoms start to pair up
     when $ T < T_{c} = 2.4 mK $ and form a superfluid \cite{he4,he3}.
     The second system is Bose-Einstein condensation in
     ultra-cold alkali atomic gases ( for a review, see \cite{rev} ).
     Recently, intensive research activities have been lavished on
     searching for excitonic superfluid in
     excitons in electron-hole semiconductor bilayers \cite{ehbl}.
     Torsional oscillator method was used to study $^{4}He $ films adsorbed on graphite
     \cite{he}, a large Non-Classical Rotational Inertial ( NCRI ) was detected
     in a narrow window of coverages $ c= 17 \sim 19 \ atoms/nm^{2} $ in the second layer
     ( $^{4}He/^{4}He$/graphite system ). The NCRI is a low temperature reduction in the
     rotational moment of inertia due to the superfluid component of the state.
     This important phenomenon was interpreted as reentrant
     "superfluid" in this narrow window  \cite{private}.
     Recently, also by using the torsional oscillator measurement, a PSU group lead by Chan
     observed a marked $ 1 \sim 2 \% $ NCRI  even in bulk solid $^{4} He $ at $ \sim 0.2 K $ \cite{chan}.
     This was interpreted as a new state of matter called supersolid which has both superfluid order
     and crystalline order \cite{chan}. However, so far, no NCRI in bulk solid $ H_{2} $
     was detected.

     At the completion of the first layer,  the $^{4}He $ atoms with the coverage
     $ c \sim 12 \ atoms/nm^{2} $ form a triangular
     lattice which is incommensurate with the underlying graphite. It is reasonable to assume the lattice
     structure of the preferred adsorption sites on the second layer form a honeycomb lattice ( Fig.1).
     When applying the Eqn.\ref{boson} to the $ ^{4}He/^{4}He $/graphite system in Fig.1, $ b^{\dagger}_{i} $ is the
    $ ^{4}He $ atom creation operator. The hopping amplitude $ t $ is determined by the trapping potential
    from the incommensurate first $ ^{4}He $ layer.  We assume the $ ^{4}He $ atoms interact with each other with the
    LJ potential. Note
    that the LJ potential is quite accurate in describing
    the long distance attractive part, but very crude in the short distance repulsive part.
    The repulsive part in real case is expected  to be softer.
    The chemical potential $ \mu $ determines
    the coverage. The filling factor $ f $ is related to the coverage $ c $ of $ ^{4}He $ with
      $ c \sim 19.55  \ atoms/nm^{2} $  corresponding to  $ f=1/2 $ where one  $ ^{4}He $ atoms occupies every two
      lattice sites of the honeycomb
      lattice ( for example, all the $ A $ sublattice sites ) to form a close packed triangular lattice
      with $ d_{AA} \sim 4.56 \AA $
      and $ d_{AB} \sim 2.63 \AA $ which is just smaller than Lennard-Jones (LJ) parameter $ \sigma(^{4}He )
      \sim 2.64 \AA $. The onsite $ U $ is very big. Because $ d_{AB} < \sigma_{ ^{4}He } < d_{AA} $, $ V_{1} $
      is positive, while $ V_{2} $ and further neighbor interactions are weakly attractive and can be neglected.
      Note that our theoretical value
      $ 19.55 \ atoms/nm^{2} $ is only $ 3 \% $ away from the experimental value $ 19 \ atoms/nm^{2} $.
      Fig.2a is very similar to the $ \rho_{s} $ versus $ c \sim f $ phase diagram
      in  $ ^{4}He/^{4}He$/graphite structure near $ c=19 \ atoms/nm^{2} $  ( see Fig. 1 in \cite{he} ).
      The dashed line is the experimental path in $^{4}He/^{4}He$/graphite at $ T=0 $.
      Ir shows that a reentrant supersolid (SS) state is a generic state
      sandwiched between the C-IC transition. The reentrant " superfluid
      " state  in the second layer of the $^{4}He $ films adsorbed on graphite could be a $ ^{4}He $
      supersolid state.
      Very interestingly, the data in the torsional oscillator experiment in \cite{he}
      do not show the characteristic form for a 2d $^{4} He $ superfluid film,
      instead it resemble that in \cite{chan} characteristic of a possible supersolid in terms of
      the gradual onset temperature of the NCRI, the unusual temperature dependence of $
      T_{SS} $ on the coverage.
      From Fig.2a, it is easy to see that it maybe difficult
      to reach the superfluid by moving along the horizontal $ r $ axis, but it is very easy to get to
      the CDW-SS state by moving along the vertical coverage axis.
      In principle, there should be a Kosterlitz-Thouless (KT) finite temperature phase transition above
      which the CDW-SS becomes
      a coexistence of the CDW state and normal fluids of interstitials.
      The effects of disorders on this finite temperature transition will be studied in  and
      compared with Fig.2 and 3 in \cite{he}.

\subsection{ $ H_2 $/Kr/graphite system }

     There are great interests in finding superfluidity in other substances.
     Hydrogen molecules are relatively light, quantum fluctuations are large at very low temperature.
     a para-H2 molecule is a boson and very similar to a He4 atom, in principle, H2 could become superfluid
     at low temperature as He4 does.  Unfortunately, unlike He4,  due to
     deeper attractive potentials, bulk H2 solidifies at low temperature
     ( $ T_{c} \sim 14 K $ ), this preempts the possible observation of the speculated superfluity.
      At sufficiently high pressure, the solid hydrogen will transform into a metallic  alkali-like crystal or
      an unusual two component ( protons and electrons ) quantum liquid at low temperatures \cite{hyd2}.
      Potential avenues to prevent $ p-H2 $ from being solidified to low
      enough temperature such that the superfluid behaviour can be observed is by decreasing the average number
      of neighbors of a $ H2 $ molecule. Monte Carlo simulations of  $ H2_{n} $ clusters with $ n=13-18 $
      showed that there is indeed a tendency for the system to be superfluid below 1K \cite{dropt}
      Indeed, Infra-red spectroscopy also provided evidences for superfludity in  $ o ( D_2)_{n} $ and $ p (H_2)_{n} $
      clusters with $ n=14-17 $ \cite{drope}. Another route is by  the reduction of dimensionality.
      Extensive theoretical and experimental work has been devoted to study H2 and D2 films in a variety of substrates.
      Path Integral Monte-Carlo (PIMC) simulations  indicated that introduction of certain impurities can
      stabilize a 2 dimensional liquid hydrogen which undergoes a Kosterlitze-Thouless (KT) transition below 1.2 K.

      In a recent experiment, neutron scattering measurements were used to characterize all the
      possible phases of $ D_2 $ coadsorbed on graphite preplated by a monolayer of Kr
      called  $ D_2 $/Kr/graphite structure \cite{honey}.
      Because two dimensional graphite has the honeycomb net structure,
      the corresponding preferred adsorption sites
      form a triangular lattice. The precoated Kr atoms will occupy the preferred adsorption sites
      and form a triangular lattice. Then the $ D_2 $ deposited on top of the Kr monolayer will sit
      on the preferred adsorption sites of the triangular lattice of the Kr monolayer to form a honeycomb lattice.
      So, the lattice geometry is very similar to Fig.1 with $ Kr $ atoms
      sitting on the triangular lattice, while the $ H_2 $ molecules
      hopping on the honeycomb lattice. The filling factor $ f $ is related to the coverage $ c $ of $ D_2 $.
      In the coverage $ c $ verse the temperature $ T $ phase diagram,
      an unusual feature is that in a small coverage range ( $ 1.20 < c < 1.25 $ ) at
      the commensurate-incommensurate (C-IC) transition, a reentrant {\em fluid} phase squeezes
      in between the C and IC phases down to $ T= 1.5 $ K which is the lowest temperature a liquidlike phase of
      $ D_{2} $ has ever been found. The filling factor $ f $ is related to the coverage $ c $ of $ H_2 $
   with $ c=1.2 $ corresponding to  $ f=1/2 $ where one $ H_2 $ molecule
   occupies every two lattice sites of the honeycomb
   lattice with $ d_{AA} \sim 3.87 \AA $ and $ d_{AB} \sim 2.24 \AA
   $ in Fig.1. Because $ d_{AB} < \sigma_{H_2} < d_{AA} $, $ V_{1} $ could also
   be positive and very big, while $ V_{2} $ and further neighbor interactions are weakly
   attractive, the above discussions on $ ^{4}He $ in subsection A can also be
   applied to this system.  Fig.2a is very similar to the $ c \sim f $ versus low $ T $ phase diagram
   in  $ H_2 $/Kr/graphite structure near $ c=1.20 $  ( Fig. 6 in \cite{honey} ).
   The dashed line is the experimental path in $ H_2 $/Kr/graphite at $ T=0 $.
   The so called $ (1 \times 1 )[\frac{1}{2}] $ commensurate phase in \cite{honey} is the CDW state where the $ D_{2} $
   atoms occupy one of the two sublattices of the underlying honeycomb lattice.
   The transition at zero temperature is a C-CDW to CDW-SS to IC-CDW transition described
   by Fig.2(a). The first transition is in the $ z=2, \nu=1/2, \eta=0 $ universality class.
   Because $ \nu < 1 $, from Harris criterion, disorders are
   relevant. The second transition is first order.
   It suggests the reentrant " fluidlike " phase could be the $ H_2 $  supersolid.
   If this is indeed the case, this may lead to the first observation of $ H_2 $ supersolid.
   Torsional oscillator experiment similar to that used in \cite{he} may be used to measure the NCRI
   of this $ H_{2} $ supersolid state. Obviously, this experiment can avoid the difficulty of using $ D_2 $
   which has large coherent neutron scattering cross section, but smaller de Boer quantum number.
   In fact, $ Kr $ may not be the best spacer to observe the $ H_{2}
   $ supersolid state. The ideal situation is to make $ d_{AB} $ is
   just slightly
   smaller than $ \sigma_{H_2} $.
   Larger atoms like $ Xe, CF_{4}, SF_{6} $ may be
   more suitable spacers, because they may increase the distance $ d_{AB} $
   in the Fig.1 just slightly smaller than  $ \sigma_{H_2} $. We suggest that $ ^{4}He
   $ supersolid should also appear in, for example,
   $^{4}He$/Kr/graphite structure.

       The reentrant lattice SS discussed in this
       section is different from the bulk  $ ^{4}He $ SS state discussed in
      \cite{qgl}, although both kinds of supersolids share many interesting common properties.
      In the former, there is a periodic substrate or spacer potential
      which breaks translational symmetries at the very beginning.
      While in the latter, the lattice results from a spontaneous
      translational symmetry breaking, so the theory developed in
      this paper on lattices is completely different from the continuum theory developed in \cite{qgl}.
      Combined with the results in \cite{qgl},
      we conclude that $ ^{4}He $ supersolid can exist both in bulk and on substrate,
      while although $ H_2 $ supersolid may not exist in the bulk,
      but it may exist on wisely chosen substrates. Ultra-cold atoms
      supersolid could also be realized in optical lattices.

\section{ Cooper-pair solid, superfluid and Cooper-pair supersolid in high temperature
         superconductors ( square lattice ) }

  Very recently \cite{cpss}, by using both angle resolved photoemission (ARPES)  and scanning tunneling
  microscopy (STM) on high temperature superconductor $ La_{2-x} Ba_{x} Cu O_{4} $ \cite{first,second},
  Valla {\sl et.al} detected a quasi-particle energy gap with $
  d $-wave symmetry even when the superconductivity is completely
  suppressed at $ x=1/8 $,  while having almost equally strong
  superconducting phases at both higher and lower dopings ( Fig.5).
  The gap turns out to reach maximum
  at $ x=1/8 $ despite $ T_{c} \rightarrow 0 $ at $ x=1/8 $. This
  fact suggests that the most strongly bound Cooper pairs at $ x=1/8
  $ are most susceptible to the charge density wave ( CDW ) ordering which suppresses $ T_{c} \rightarrow 0 $.
  In this section, I propose
  that the formation of a stripe Cooper pair supersolid may be able to explain
  the very unusual phase diagram of
  $ La_{2-x} Ba_{x} Cu O_{4} $ at and near doping level $ x=1/8 $.

  The fact that the STM and ARPES measurements in \cite{cpss} still detected an energy gap
  at Fermi surface with $ d_{x^2-y^2} $ wave symmetry  suggests that there are tightly bound
  Cooper pairs in this cuprate. We can treat these Cooper pairs as bosons, therefore many
  important results achieved in previous sections on interacting bosons hopping on lattices maybe
  applied to this cuprate. One important new  feature for tightly bound
  Cooper pairs is that they carry charges $ 2e $, therefore interact
  with bare long-range Coulomb interaction ( in high $ T_{c} $ cuprates, by
  taking  the penetration depth $ \lambda \rightarrow \infty $,
  one can neglect the very weak Meissner effect )  \cite{yeboson}.
  I propose the following physical picture:
  at exactly $ x=1/8 $, the filling factors of the bosons on the square lattice is $ f=\frac{1-x}{2}=7/16 $,
  the system is a CDW ( more specifically, stripe ) of Copper pairs with 4 sites per unit cell, so $ T_{c}=0 $.
  Then we need $ q=16 $
  dual vortices to describe the stripe CDW in the dual vortex picture. It
  was explicitly pointed out in section III that special
  care is needed to choose the correct saddle point of the dual gauge field
  $ A_{\mu} $ to describe this stripe CDW state.
  {\em Slightly away from $ x=1/8 $, on both sides,
  the groundstate is a  Stripe Cooper pair supersolid ( CP-SS ) ! }
  When $ x>1/8 $ $  ( x <1/8 ) $, it is a hole ( electron ) doped stripe CP-SS.
  Indeed, as shown by Quantum Monte-Carlo simulation \cite{add},
  only striped supersolid is stable in a square lattice for hard-core bosons.
  The quantum phase transition from the stripe Cooper pair solid to
  the stripe CP-SS driven by the doping is in the same universality class as that from a
  Mott insulator to a superfluid, therefore have exact exponents $ z=2, \nu=1/2, \eta=0 $
  ( with possible logarithmic corrections ).
  The resulting stripe CP-SS has {\sl the same lattice symmetry
  breaking patterns} as the stripe Cooper pair solid with the superfluid density
  scaling as $ \rho_{s} \sim |x-1/8 |^{(d+z-2)\nu }=| x-1/8 |= |\delta x | $ with
  a possible logarithmic correction.
  As pointed out in section III, the superfluid density is
  anisotropic with $ \rho_{s} $ along the stripe is larger than that normal to
  the stripe, but both scale in the same way with different
  coefficients. From Uemura relation, we conclude that $ T_{c} \sim |\delta x | $ ( in fact, it scales as the smaller
  $ \rho_{s} $ in the Stripe CP-SS ).
  This  result very naturally explains why $ T_{c} $ indeed looks linear in
  $  |\delta x | $ and  the phase diagram is nearly symmetric near $ x = 1/8 $ found in \cite{cpss}.

\begin{figure}
\includegraphics[width=6cm]{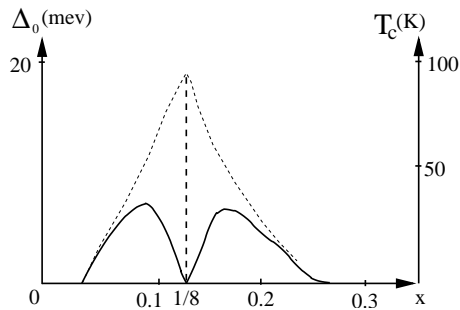}
\caption{ The dashed line is the energy gap $ \Delta_{0} $ which
reaches minimum at $ x=1/8 $, the solid line is the superconducting
transition temperature $ T_{c} $ which has two domes and is zero at
$ x=1/8 $. }
 \label{fig6}
\end{figure}

  The above picture only involves the sector of the tightly bound Copper pair. Of
  course, understanding the ground state of the Cooper-pair sector is very important on its own
  and is also the starting point to incorporate quasi-particles \cite{s0,random,s1} and
  spin excitations \cite{third} into the picture.  As shown in section IV, a
  valence-bond supersolid ( VB-SS ) can be stabilized if there is
  a considerable ring exchange interaction. An interesting
  question to address is if valence-bond Cooper-pair supersolid can be
  realized in some of these cuprates.

\section{ Conclusions }

   By using the DVM, we studied superfluid, solid and
   supersolid and quantum phase transitions of the extended boson
   Hubbard model near half filling on bipartite optical lattices such
   as honeycomb and square lattice and  mapped out the global phase
   diagram at $ T=0 $ in a unified scheme.
   We identified boson density and boson kinetic energy operators in
   terms of the dual vortex fields
   to characterize symmetry breaking patterns in the insulating states and supersolid states.
   In the DVM, starting from the featureless superfluid state where the
   the average value of the dual gauge field through a dual plaquette  is taken to be uniform
   and equal to the boson density $ f=p/q $, its fluctuation is
   coupled to $ q $ dual vortex order parameters and is gapless, one study
   all the possible symmetry breaking patterns by condensing the $ q
   $ vortex order parameters.
   We first study the transition driven by the ratio of the kinetic
   energy over the potential energy at the commensurate fillings $
   f=p/q $ along the horizonal axis in Fig.2-4. In the Ising limit, we found that the saddle point of
   the dual gauge fields should be chosen in a self consistent way
   in the CDW side, this is in sharp contrast to the more familiar $
   f=n $ case where the self-consistency condition is automatically
   guaranteed.  Due to this change of saddle point structure of the dual gauge fields on the both sides, the SF
   to the CDW transition is always  a strong first order one.
   In the Easy-plane limit, we found that the self-consistency condition is automatically
   guaranteed, the SF to the VBS transition is a weak first order one.
   Then we study the transition driven by chemical potential slight away from the commensurate fillings $
   f=p/q $ along the vertical axis in Fig.2-4.
   We found that in the insulating side, the transition at zero temperature
   driven by the chemical potential must be
   a C-CDW ( or C-VBS ) at half filling
   to a narrow window of CDW- ( VB-) supersolid,
   then to a IC-CDW ( IC-VBS ) transition in the Ising ( easy-plane ) limit.
   The valence bond supersolid is a novel kind of supersolid first proposed in this paper.
   Although the density fluctuation at {\em  any site} is very large
   indicating its superfluid nature, the boson  {\em kinetic energies }
   {\em on bonds} between two sites are given and break the lattice translational symmetry
   indicating its valence bound nature.
   The first transition is in the  same universality class as that from a Mott insulator to
   a superfluid driven by a chemical potential, therefore have exact exponents $ z=2,
   \nu=1/2, \eta=0 $ with a logarithmic correction. The second is a 1st order transition.
   The results achieved in this letter could guide QMC
   simulations to search for all these phases and confirm the
   universality class of the transitions. This VB-SS should be stable
   in both hard core and sift core limit if there is a sufficient
   strong ring exchange term in Eqn.\ref{boson}. It will be
   extremely interesting to search this novel kind of supersolid in
   a specific microscopic model by QMC simulation.

     In the second part of this paper, we study the applications of the results achieved in the first part by the DVM to
     3 very important ultra-cold atomic and condensed matter experimental systems. The EBHM in honeycomb and square lattices could be realized in
     ultracold atoms loaded on optical lattices.
     So the results achieved in this paper may have direct impacts on the atomic experiments  in optical lattices.
     Then we applied the results to two condensed matter
     experimental systems: (1)  adatom adsorption on different substrates
     such as the adsorption in the second layer of  $ ^{4}He $ adsorbed on
             graphite and Hydrogen adsorbed on Krypton-preplated
             graphite in honeycomb lattice, we find that a reentrant
             $ ^{4}He $ supersolid in the second layer on graphite may be
             responsible for the NCRI detected in the torsional oscillator
             experiments in \cite{he} and a reentrant
             $ H_2 $ supersolid  at $ T=0 $ maybe responsible for the reentrant
             liquid state squeezed between C and IC phases detected
             by coherent neutron scattering.
             We propose that a judicious choice of substrate may also lead
             to an occurrence of hydrogen lattice supersolid. (2)
             superconducting phase diagram  near $ x=1/8 $ in high temperature superconductor
             $ La_{2-x} Ba_{x} Cu O_{4} $ in square lattice. We
             conclude that there is a stripe CDW at $ 1/8 $ which
             suppresses $ T_{c} $ to zero. There are
             hole and electron doped Cooper pair supersolids on both sides of $ x=1/8 $ which has the
             same lattice symmetry breaking pattern as the stripe
             CDW at $ x=1/8 $ and with $ \rho_{s} \sim T_{c}  \sim
             |x-1/8 | $. Although the SS in lattice models is different from that in a continuous systems,
     the results achieved in this paper on lattice supersolids may still shed some lights on the possible
     microscopic mechanism and phenomenological Ginsburg-Landau
     theory of the possible $ ^{4}He $ supersolids \cite{qgl}.

    I thank Milton Cole for pointing out
    the experiment in Ref.\cite{honey} to me. I also like to thank A. Millis
    for pointing out Ref.\cite{cpss} to me and helpful discussions.
    This research at KITP was supported in part by the NSF
    under Grant No. PHY99-07949 at KITP-C
    by the Project of Knowledge Innovation Program (PKIP) of Chinese Academy of Sciences.
    I also thank C.P. Sun for hospitality during my visit at
    KITP-C.

\appendix

\section{ Duality at integer fillings $ f=n $, the role of the dual gauge field and the self-consistency
          condition  }

     It is instructive to review the direct boson picture and the duality
     transformation to the dual vortex picture in this simplest case with integer filling $ f=n $ \cite{dualint}.
     The contents of this appendix are not new. However, we stress the
     self-consistence check on the average value of the dual
     magnetic field on both sides of SF and Mott insulator.
     We also explicitly spell out the physical significance of the fluctuations
     of the dual gauge field on both sides. These clarifications are
     very helpful to motivate the self-consistence condition
     on the average value of the dual
     magnetic fields in the  translational symmetry breaking insulating sides, especially in the
     CDW side at $ q \geq 2 $ cases first discovered in \cite{univ}. Although this
     self-consistence condition is automatically satisfied at $
     f=n $ case, they become a  non-trivial constraint on a
     self-consistent theory at $ q \geq 2 $ cases. In a recent
     unpublished note \cite{kagome}, the author found this self-consistent
     condition is even more non-trivial and important in frustrated lattices such as
     triangular and Kagome lattices than the bipartite lattices
     discussed in this paper. As shown in the
     main text, in both the CDW side ( using limit ) and the valence
     bond ( easy plane limit ), at and slightly away from the
     commensurate filling $ f=p/q $ with $ q \geq 2 $ cases, the effective actions can be
     mapped to Eqn.\ref{vortex1} and Eqn.\ref{vortex2}.

     In the direct picture, it is convenient to start from the Mott insulating side and look at its low
     energy excitations,  it is easy to see that the creation of a particle by $
     \phi^{\dagger} $ is always accompanied by a creation of a hole
     $ \phi $ due to the particle hole symmetry at $ f=n $, so
     the Ginzburg-Landau theory to describe the Mott to superfluid
     transition in terms of the boson order parameter $ \phi $ is given by the well know
     $ 2+1 $ dimensional relativistic complex scalar theory:
\begin{equation}
   {\cal S}_{b}= \int d^{2}r d \tau [ | \partial_{\tau} \phi |^{2} +
   |\nabla \phi |^{2} + r |\phi |^{2} + u |\phi |^{4} +\cdots  ]
\label{boson1}
\end{equation}
    In the Mott state $ r > 0 $, so $  \langle \phi \rangle =0 $, there is a Mott gap. In the
    superfluid state  $ r < 0 $, so $  \langle \phi \rangle \neq 0 $. Due to the
    symmetry breaking in the SF state, there is a gapless goldstone
    mode given by the phase fluctuation of $ \phi $.

    In the dual vortex picture,  it is convenient to start from the Superfluid side and look at its low
    energy excitations. There are two kinds of low energy
    excitations. The first is just the gapless Goldstone mode in the phase of $ \phi $ which
    is given by a dual gauge field fluctuation $ A_{\mu} $. The second
    is the  topological vortex excitation in the phase winding of $
    \psi $. Obviously, the number of vortex $ \psi^{\dagger} $ is equal to the number
    of anti-vortex $ \psi $, so the Ginzburg-Landau theory to describe the
    superfluid to the Mott  transition in terms of the dual vortex order parameter
    $ \psi $ is given by the $ 2+1 $ dimensional scalar electrodynamics:
\begin{eqnarray}
   {\cal S}_{d}= \int d^{2}r d \tau  & [ & | ( \partial_{\mu} -i A_{\mu} ) \psi |^{2} +
   + r_{d} |\psi |^{2} + u_{d} |\psi |^{4} +\cdots
   \nonumber  \\
   & + &  \frac{1}{4}
   ( \epsilon_{\mu \nu \lambda} \partial_{\nu} A_{\lambda} )^{2} ]
\label{vortex1}
\end{eqnarray}
    In the superfluid state $ r_{d} > 0 $, so $  \langle \psi \rangle =0 $, there is a gapless fluctuation
    given by the dual gauge field $ A_{mu} $. Integrating out the gauge field fluctuation
    $ A_{\mu} $ will lead to a logarithmic interaction between the
    vortices. In the Mott insulating state  $ r_{d} < 0 $, so $  \langle \psi \rangle \neq 0 $. Due to the
    "symmetry" breaking in the Mott insulating state, the dual gauge field $ A_{\mu} $ acquires
    a mass due to Higgs mechanism, so there is a Mott gap in the
    Mott phase. The vortex action Eqn.\ref{vortex1} is dual to the
    boson action Eqn.\ref{boson1}. Both actions lead to equivalent
    description of the SF to Mott transition which is in the $ 3D XY $ universality class
    and the same excitation spectra  in both phases. Indeed, this universality class has been confirmed
    by QMC in the first reference in \cite{dualint} from both the direct boson
    action Eqn.\ref{boson1} and the dual vortex
    action Eqn.\ref{vortex1}.

    In the direct boson picture, a vortex is a singularity in the boson wavefunction, so a boson
    wavefunction acquires a $ 2 \pi $ phase when in encircles a vortex.
    In the dual vortex picture, a boson is a singularity in the vortex wavefunction, so a
    vortex  wavefunction acquires a $ 2 \pi $ phase when in encircles a
    boson. So the average strength of the dual magnetic field of the gauge field $ A_{\mu}
    $ through a dual plaquette is equal to the boson density $ f=n
    $, because $ 2 \pi n $ is equivalent to $ 0 $, so the average value can be simply taken to be zero.
    It is important to stress that the average density of bosons is the
    same in both the SF and the Mott insulating side, namely, it
    takes the integer $ n $ on both sides, so the average strength of the dual magnetic field can be
    taken as zero on both side, then the fluctuations in the dual
    gauge field reflects the fluctuations of the boson density.
    As said in the last paragraph, the gauge field is gapless in the
    SF phase, so the boson density fluctuation in the SF is very
    large, this is expected, because the SF is a phase ordered
    state, so it has a large density fluctuation. However, in the Mott
    phase, the gauge filed is massive, so the density fluctuation is
    suppressed in the Mott phase. This is expected also, because the
    Mott phase is a density ordered phase, so has very little
    density fluctuations. So by looking at the behaviors of the dual
    gauge field on both sides, one can distinguish the properties of the two phases.
    In short, although the average value of the dual gauge field is
    the same on both the SF and the Mott insulating side, its
    fluctuations are completely different which, in turn, is
    determined by the average value of the vortex order parameter $
    \psi $ in Eqn.\ref{vortex1}.

    When slightly away from $ f=n $ which is at a in-commensurate density, there is no particle-hole
    symmetry anymore, so there should be a first order imaginary time derivative,
    Eqn.\ref{boson1} becomes:
\begin{equation}
   {\cal S}_{ic-b}= \int d^{2}r d \tau [ \phi^{\dagger} \partial_{\tau} \phi +
    |\nabla \phi |^{2} + r |\phi |^{2} + u |\phi |^{4} +\cdots  ]
\label{boson2}
\end{equation}
    where we have dropped the second derivative term $ | \partial_{\tau} \phi
    |^{2} $ which is less important than the linear derivative term.
    If there is only a onsite interaction in
    Eqn.\ref{boson}, then away from $ f=1 $, the system is always in the superfluid state.
    The transition from the Mott to the SF transition driven by the chemical potential  has the critical
    exponents $ z=2, \nu=1/2, \eta=0 $ with  a logarithmic correction.

    In the dual vortex picture, the linear time derivative term corresponds
    to adding a small mean dual magnetic field  $ \delta f= f-n $ to Eqn.\ref{vortex1}:
\begin{eqnarray}
   {\cal S}_{ic-d}= \int d^{2}r d \tau  & [ & | ( \partial_{\mu} -i A_{\mu} ) \psi |^{2} +
   + r_{d} |\psi |^{2} + u_{d} |\psi |^{4} +\cdots
   \nonumber  \\
   & + &  \frac{1}{4}
   ( \epsilon_{\mu \nu \lambda} \partial_{\nu} A_{\lambda} - 2 \pi  \delta f \delta_{\mu \tau}  )^{2} ]
\label{vortex2}
\end{eqnarray}
     This action is essentially the same as the Ginzburg-Landau
     model for a superconductor in a external magnetic field if we
     identify the $ \tau  $ direction as the $ \hat{z} $ direction
     along which the magnetic field $ \delta f $ is applied.
     For a type II superconductor, its phase diagram is shown in Fig.2a.


\begin{thebibliography}{99}
\bibitem{equal} The discussions  in the following equally apply to the striped solid with $
(\pi,0) $ or $ (0,\pi) $ order with the exception of the anisotropy
in the superfluid density. The anisotropy is irrelevant to the
universality class.
\bibitem{direct} In the direct lattice picture, we can simply renormalize away the
sublattice  $ A $ where the bosons sit and focus on the effective
boson hopping on the sublattice $ B $ whose dual vortex action is
given by Eqn.\ref{is}. If $ \delta f $ is replaced by $ -\delta f $,
then the $ b $ index should be replaced by the $ a $ index, the
vortices in the phase winding of $ \psi_{a} $ should be interpreted
as the vacancy number.
\bibitem{chem} In principle, the $ \delta f $ should be the chemical potential.
\end{thebibliography}
\end{document}